\documentclass[aps,
twocolumn,
superscriptaddress,
amssymb,
amsmath,
prb,
floatfix,
longbibliography]{revtex4-2}
\usepackage{graphicx }
\usepackage[]{hyperref}
\usepackage{dcolumn}

\hypersetup{colorlinks=true,linkcolor=blue,citecolor=blue,urlcolor=blue,pdfpagemode=UseNone}

\newcommand{\slrr}      {$T_1^{-1}$}

\begin{document}
\title{Absence of strong magnetic fluctuations or interactions in the normal state of LaNiGa$_2$}

\author{P. Sherpa}
    \affiliation{Department of Physics and Astronomy, University of California, Davis, California
95616, USA}

\author{I. Vinograd}
    \affiliation{Department of Physics and Astronomy, University of California, Davis, California
95616, USA}

\author{Y. Shi}
    \affiliation{Department of Physics and Astronomy, University of California, Davis, California
95616, USA}
\author{S. A. Sreedhar}
    \affiliation{Department of Physics and Astronomy, University of California, Davis, California
95616, USA}
\author{C. Chaffey}
    \affiliation{Department of Physics and Astronomy, University of California, Davis, California
95616, USA}
\author{T. Kissikov}
    \affiliation{Department of Physics and Astronomy, University of California, Davis, California
95616, USA}

\author{M.-C. Jung}
    \affiliation{Department of Physics, Arizona State University, Tempe, AZ 85287, USA}

\author{A. S. Bontana}
    \affiliation{Department of Physics, Arizona State University, Tempe, AZ 85287, USA}

\author{A. P. Dioguardi}
    \affiliation{Los Alamos National Laboratory, Los Alamos, New Mexico 87545, USA}

\author{R. Yamamoto}
    \affiliation{Los Alamos National Laboratory, Los Alamos, New Mexico 87545, USA}

\author{M. Hirata}
\affiliation{Los Alamos National Laboratory, Los Alamos, New Mexico 87545, USA}

\author{G. Conti}
    \affiliation{Advanced Light Source, Lawrence Berkeley National Laboratory, Berkeley, California, 94720, USA}

\author{S. Nemsak}
    \affiliation{Advanced Light Source, Lawrence Berkeley National Laboratory, Berkeley, California, 94720, USA}

\author{J. R. Badger}
    \affiliation{Department of Chemistry, University of California, Davis, California
95616, USA}
\author{P. Klavins}
    \affiliation{Department of Physics and Astronomy, University of California, Davis, California
95616, USA}
\author{I. Vishik}
    \affiliation{Department of Physics and Astronomy, University of California, Davis, California
95616, USA}
\author{V. Taufour}
    \affiliation{Department of Physics and Astronomy, University of California, Davis, California
95616, USA}
\author{N. J. Curro}
    \affiliation{Department of Physics and Astronomy, University of California, Davis, California
95616, USA}
\date{\today}
\begin{abstract}

We present nuclear magnetic (NMR) and qudrupole (NQR) resonance and magnetization data in the normal state of the topological crystalline superconductor LaNiGa$_2$. We find no evidence of magnetic fluctuations or enhanced paramagnetism. These results suggest that the time-reversal symmetry breaking previously reported in the superconducting state of this material is not driven by strong electron correlations.
\end{abstract}

\maketitle

\section{Introduction}

The emergence of unconventional superconductivity is generally accepted to be a consequence of electron-electron interactions in materials that usually exhibit strong magnetic correlations in the normal state \cite{MonthouxPinesReview}. These correlations can also play an important role in the behavior of the class of unconventional superconductors that break time-reversal symmetry (TRS) in the superconducting state.  This property reveals important information about the nature of the superconducting condensate, such as triplet pairing, or if there are multiple components of the superconducting order parameter \cite{UPt3spintripletNMR,Mineev1999,huiqui,Kallin2016}.
The vast majority of superconductors do not exhibit TRS breaking, however those that do may have non-trivial topological properties that could support Majorana zero modes, which potentially could be exploited as dissipationless qubits for quantum computing \cite{Sato2017a,TopoQuantReview}. Determining the presence and origin of TRS breaking in the superconducting state is challenging, because the associated magnetic field is typically very small and is usually detected only via muon spin relaxation ($\mu$SR) \cite{Luke1998,BlundellDeRenziTRSB2021} or polar Kerr effects \cite{polarKerr2006}.

The intermetallic superconductor LaNiGa$_2$ has recently attracted attention because $\mu$SR experiments in this material uncovered TRS breaking in the superconducting state below $T_c = 2.1$ K \cite{Hillier2012}. This material has a similar stoichiometry as LaNiC$_2$ \cite{LaNiC2TRS2009}, which also exhibits TRS breaking due to a combination of spin-orbit coupling and a non-centrosymmetric structure  \cite{GorkovNonCentroSymm}. However, LaNiGa$_2$ is centrosymmetric and recent penetration depth, specific heat, {and $\mu$SR} measurements have revealed multiple, nodeless gaps~\cite{Weng2016,Quintanilla2020,Sundar2023}.  A recent single crystal study revealed that LaNiGa$_2$ actually has a non-symmorphic crystal structure that gives rise to a non-trivial band topology \cite{Badger2022} with band-degeneracies at the Fermi level. This electronic structure can support interband pairing and a superconducting order parameter that can be antisymmetric in the band channel, allowing for fully-gapped equal-spin pairs.



An important open question is what drives the imbalance between two equal-spin gaps resulting in the time-reversal symmetry breaking that was observed below $T_c$~\cite{Hillier2012}. Non-unitary multiorbital superconductivity may arise from competing interactions~\cite{Wolf2022}, and spin fluctuations are generally present in the normal state of unconventional superconductors~\cite{Curro2005,MonthouxPinesReview}.  It is therefore important to investigate the strength of electron correlations that may be present in the normal state of LaNiGa$_2$. Here we report nuclear magnetic resonance (NMR), nuclear quadrupole resonance (NQR), bulk magnetization, and X-ray photoelectron spectroscopy (XPS) measurements, as well as density-functional theory (DFT) calculations, that reveal the absence of any significant spin fluctuations or Stoner enhancement, suggesting that electron correlation effects in this material are weak and therefore  unlikely to play a role in the unusual superconducting pairing.

\section{Methods}

\begin{figure}
\centering
\includegraphics[width = \linewidth]{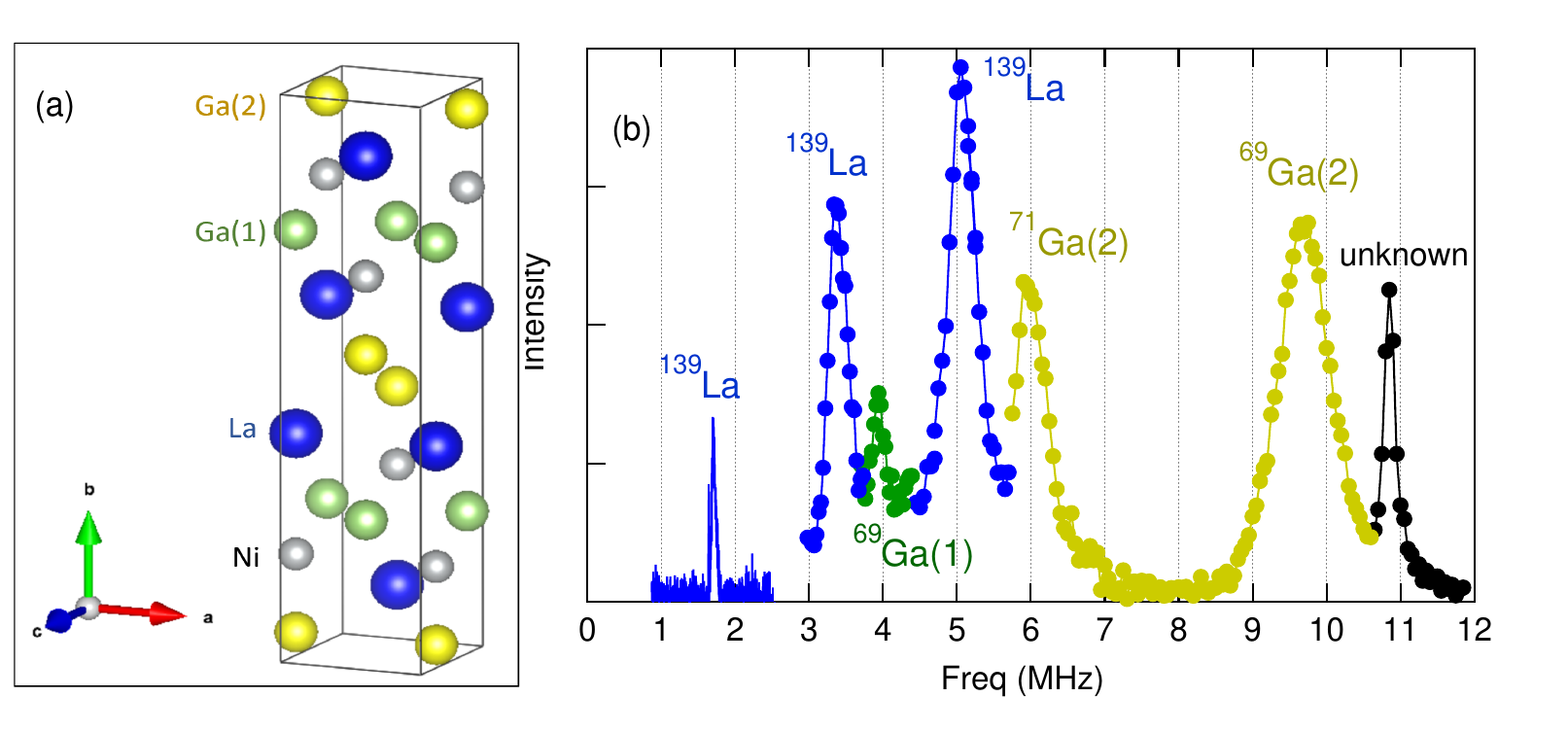}
\caption{(a) Unit cell of LaNiGa$_2$. (b) NQR spectrum measured at 4 K.  The $^{71}$Ga(1) resonance near 2.5 MHz is not shown.
}
\label{fig:UnitCell}
\end{figure}

Single crystals of LaNiGa$_2$ were grown via flux methods as described in \cite{Badger2022}. This material has one La site and two crystallographically distinct Ga sites [dubbed Ga(1) and Ga(2) hereafter], as illustrated in Fig. \ref{fig:UnitCell}(a).  Magnetization measurements were performed with a Magnetic Property Measurement System (MPMS, Quantum Design) in the temperature range of 2\,K to 300\,K. Because the magnetic susceptibility is relatively small, we prepared a mosaic of co-aligned single crystals, allowing for a larger signal.

XPS measurements were performed using a lab-based XPS setup (Kratos Axis Supra). The Ga $2p_{3/2}$ core levels were obtained using an Al K-$\alpha$ source and Ag L-$\alpha$ on the single crystals at room temperature.

\begin{table}[b]
\caption{\label{tab:NMRpars}NMR parameters for the three isotopes measured in LaNiGa$_2$.}
\begin{ruledtabular}
    \begin{tabular}{llll}
        {isotope} & $^{139}${La} & $^{69}${Ga} & $^{71}${Ga} \\ \hline
        abundance & 99.1\% & 60\% & 40\% \\
        $I$  & 7/2 & 3/2 & 3/2 \\
        $\gamma/2\pi$ (MHz/T) & 6.0146 & 10.219 & 12.985 \\
        $Q$ (barn) & 0.21 & 0.178 & 0.112 \\
\end{tabular}
\end{ruledtabular}
\end{table}

For the NMR measurements, three single crystals were aligned to make a mosaic with dimensions $1.3\times0.5\times 0.5$ mm$^3$, secured in an coil, and placed in an external field in a cryostat. The resonance frequencies are determined by the Hamiltonian:
\begin{equation}
    \mathcal{H} = \gamma \hbar \mathbf{\hat{I}}\cdot(1 + \mathbf{K})\cdot\mathbf{H}_0+ \frac{h\nu_{zz}}{6}\left[3\hat{I}_z^2 - \hat{I}^2 + \eta(\hat{I}_x^2 - \hat{I}_y^2)\right]
    \label{eqn:H}
\end{equation}
where $\gamma$ is the gyromagnetic ratio, $h$ ($\hbar$) is the Planck (reduced Planck) constant, $\mathbf{\hat{I}}$ is the nuclear spin angular momentum operator, $\mathbf{H}_0$ is the external magnetic field vector, $\mathbf{K}$ is the NMR shift tensor, $\eta =(\nu_{xx} - \nu_{yy})/\nu_{zz}$ is the asymmetry parameter, and $\nu_{\alpha\alpha} = 3eQV_{\alpha\alpha}/2I(2I-1)h$ are the principal values of the electric field gradient (EFG) tensor, $V_{\alpha\beta}$, (where $\alpha$ and $\beta$ stand for one of the three directions of the principal axes of the EFG tensor), $I$ is the nuclear spin quantum number, and $Q$ is the nuclear quadrupolar moment. The NMR parameters for each isotope is given in Table \ref{tab:NMRpars}. NMR spectra were measured by integrating the echo intensity as a function of frequency in either a field of $\mu_{0}H_0=11.7286$ T or 7.0 T at 5 K for fields both parallel and perpendicular to the $b$-axis. The Knight shift and EFG components, $K_{\alpha\alpha}$ and $\nu_{\alpha\alpha}$, were determined by fitting the full spectra to exact diagonalization results for Eq. \ref{eqn:H} for various orientations of $\mathbf{H}_0$. NQR spectra were acquired at zero applied field at 4 K by integrating the Fourier transform of the echo intensity as a function
of frequency.

The spin-lattice relaxation rate, \slrr, was measured by NMR using the inversion recovery method at the central transition ($I_z = +1/2 \leftrightarrow -1/2$) of $^{139}$La and $^{69}$Ga(1) sites as a function of temperature in a magnetic field of 7 T.   The recovery of nuclear magnetization after inversion for the $^{139}$La site was fitted to the standard expression for a nuclear spin $I = 7/2$ system:
$M(t)\!=\!M_0\left(1-2f\sum_n A_n e^{-\alpha_n t/T_1}\right)$,
where $M_0$ is the equilibrium nuclear magnetization, $f$ is the inversion fraction, $A_1 = 1225/1716$, $A_2 = 75/364$, $A_3 = 3/44$, $A_4= 1/84$, $\alpha_1 = 28$, $\alpha_2 = 15$, $\alpha_3  = 6$, and $\alpha_4 = 1$.    For the $^{69}$Ga(1) site with a nuclear spin $I = 3/2$, the recovery was fitted using $A_1 = 9/10$, $A_2 = 1/10$,  $\alpha_1 = 6$, and $\alpha_2 = 1$. There was no evidence of any stretched relaxation, any signal wipeout, or quadrupolar relaxation.

{We performed density-functional theory (DFT)-based calculations for LaNiGa$_2$ using the all-electron full-potential code, Wien2k~\cite{wien2k}. The exchange-correlation functional used was the Perdew-Burke-Erznerhof version of the generalized gradient approximation (PBE-GGA)~\cite{pbe}. The number of plane
waves was limited by a cut-off set by $R_{mt}$$K_{max}$ = 7 and the muffin-tin radii used were 2.5 a.u. for La, 2.32 a.u. for Ni, and 2.20 a.u. for Ga the atoms. In order to obtain the EFG tensors, we used a very fine k-mesh of 34 $\times$ 34 $\times$ 33 in the irreducible Brillouin zone. }

\section{Results}

\subsection{Stoner enhancement factor}

\begin{figure}[]
	\centering
	\includegraphics[width=\linewidth]{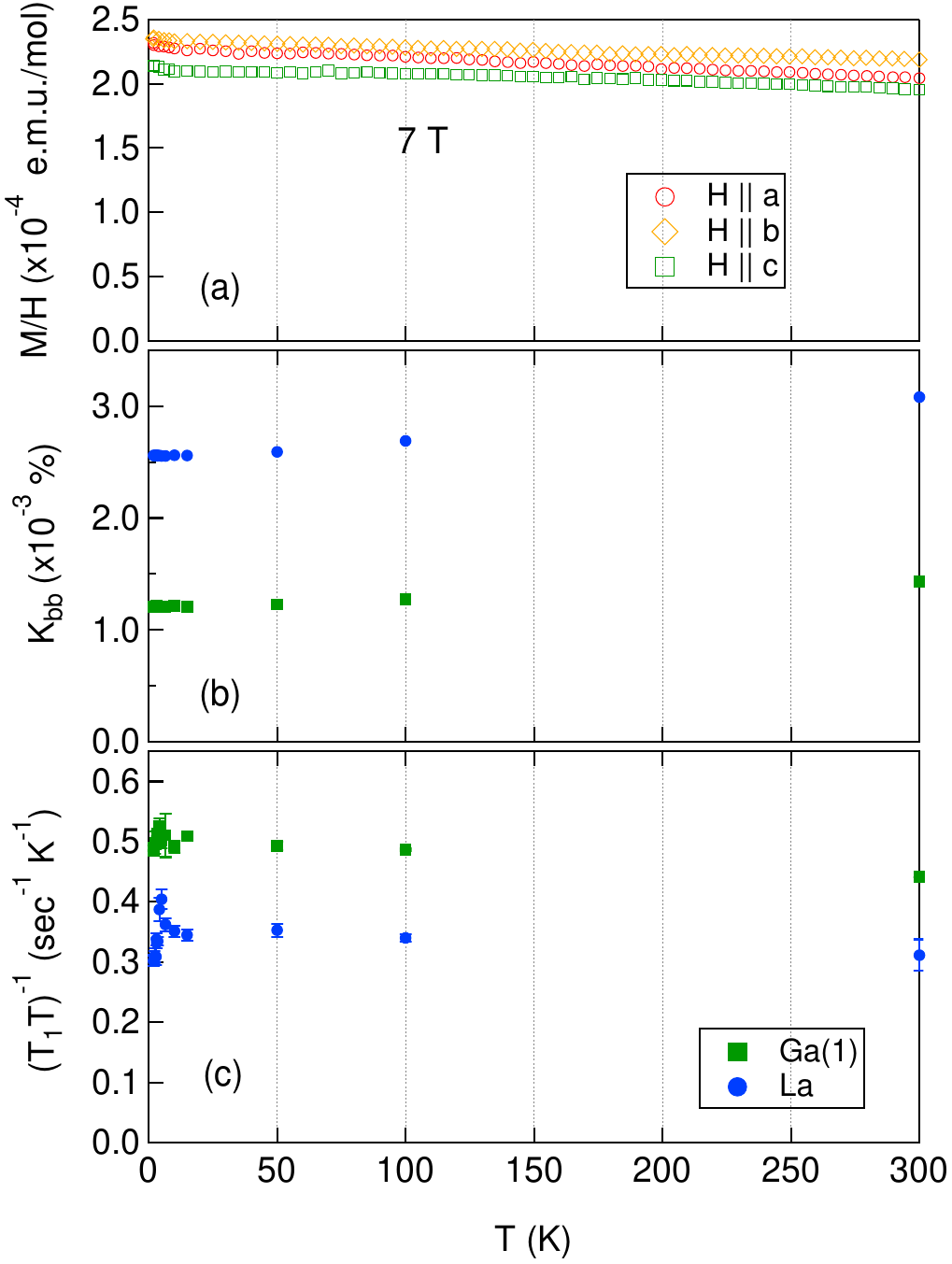}
	\caption{(a) Susceptibility as a function of temperature of mosaic LaNiGa$_2$ crystals with the magnetic field along the $a$, $b$ and $c$ axes. (b) The Knight shift $K$ and (c) the spin-lattice relaxation rate divided by temperature $(T_1T)^{-1}$ plotted against temperature for both the $^{139}$La and $^{69}$Ga(1) sites for the field along the $b$ axis.
	\label{susceptibility}}
\end{figure}

Figure~\ref{susceptibility}(a) shows the DC magnetic susceptibility of a mosaic of LaNiGa$_2$ single crystals with an applied magnetic field of 7\,T along the $a$, $b$ and $c$ axes. The susceptibility appears to be almost temperature-independent, suggesting Pauli paramagnetic behavior. By averaging the susceptibility values across the entire temperature range, we obtain susceptibilities of $2.17\times10^{-4}$, $2.26\times10^{-4}$ and $2.05\times10^{-4}$ e.m.u./mol along the $a$, $b$ and $c$ axes, respectively.

\begin{figure*}[!t]
\centering
\includegraphics[width = \linewidth]{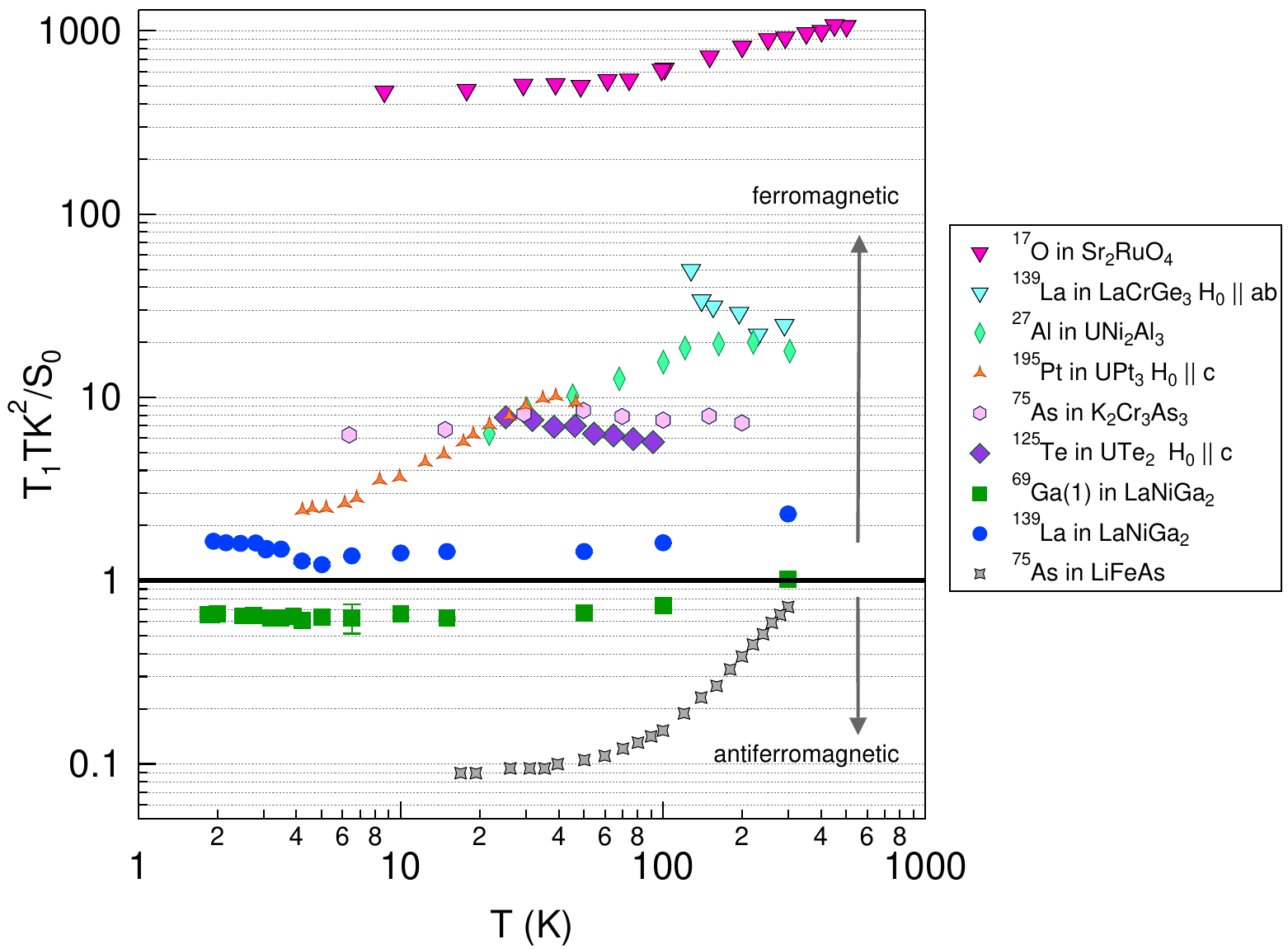}
\caption{The Korringa ratio, $T_1TK^2/S_0$, versus temperature for LaNiGa$_2$ (for field along the $b$ direction) compared with data for LiFeAs \cite{LiFeAsNMR2010},
UTe$_2$ \cite{Tokunaga2019}, UNi$_2$Al$_3$ \cite{UNi2Al3KnightShift2002}, LaCrGe$_3$ \cite{rana2019magnetic}, K$_2$Cr$_3$As$_3$ \cite{Yang2021a}, UPt$_3$ \cite{HalperinUPt31993}, and Sr$_2$RuO$_4$ \cite{Imai1998}. 
}
\label{fig:Korringa}
\end{figure*}

In general, the measured temperature-independent susceptibility $\chi_{meas}$ consists of two components: the paramagnetic contribution from conduction electrons, $\chi_P$, and the diamagnetic contribution, $\chi_D$, from the atomic cores. According to DFT calculations, the Ni 3$d$ band is filled~\cite{Quan2022Nonsymmorphic}. We therefore can estimate the Langevin diamagnetic susceptibility of Ni$^{3d^{10}}$ by extrapolating from that of other ions with the same 3$d$$^{}$ electron configuration, such as Cu$^{+}$, Zn$^{2+}$, Ga$^{3+}$, and Ge$^{4+}$~\cite{Mendelsohn1970}.  This gives a value of  $\chi_{D,\text{Ni}^{3d^{10}}}=-19\times10^{-6}$\,e.m.u./mol. For the La site we use $\chi_{D,\text{La}^{3+}} = -20\times10^{-6}$\,e.m.u./mol~\cite{Bain2008Diamagnetic}.

In order to determine the diamagnetic contribution from the Ga, we first performed XPS to determine the Ga electronic configuration. Figure~\ref{fig:XPS} in the appendix shows sub-peaks corresponding to neutral Ga(0) (BE=1116.4\,eV)~\cite{Schon1973JES,Wagner1975FDCS} and Ga(III) (BE= 1117.8\,eV)~\cite{Schon1973JES,Wagner1975FDCS}. The Ga(III) peak shows significant attenuation between freshly cleaved (Fig.~\ref{fig:XPS}(b)) and air-exposed samples (Fig.~\ref{fig:XPS}(a)) with the binding energy matching well with literature values for Ga$_2$O$_3$~\cite{Schon1973JES,Wagner1975FDCS}. This observation agrees with the species being a surface oxide as seen by changing to a Ag L-$\alpha$ source (Fig.~\ref{fig:XPS}(c)) wherein the relative ratio of the peaks changes to be more bulk dominated~\cite{Seah1979SIA}. We conclude that the bulk oxidation state of Ga is neutral, therefore  $\chi_{D,\text{Ga}^0}=-32\times10^{-6}$\,e.m.u./mol~\cite{Mendelsohn1970}. This yields a total diamagnetic susceptibility $\chi_D = -10.3\times10^{-5}$\,e.m.u./mol for LaNiGa$_2$.

The Pauli susceptibility of a free electron gas
is given by $\chi_P = \left({3\mu_0\mu_B^2 10^6}/{4\pi^3k_B^2}\right)\gamma_S$, where $\gamma_S$ is the Sommerfeld coefficient.
Using $\gamma_S = 14.1$\,mJ$\cdot$mol$^{-1}$$\cdot$K$^{-2}$~\cite{Badger2022}, we obtain $\chi_P = 1.93\times10^{-4}$\,e.m.u./mol. Using the values for $\chi_P$ and $\chi_D$, we can now extract the Stoner enhancement factor, $Z$,  from the measured susceptibility via the relation:
\begin{equation}
    \frac{1}{1-Z}=\frac{\chi_{meas}-\chi_D}{\chi_P}.
\end{equation}
We find $Z= 0.40$, 0.41, and 0.37 along the $a$, $b$, and $c$ axes respectively. These $Z$ values are smaller than the Stoner limit ($Z=1$) and comparable with the estimated value of copper, $Z = 0.26$ (using $\gamma_S$=0.505 mJ$\cdot$mol$^{-1}$$\cdot$K$^{-2}$ \cite{Kittel2004} and $\chi_0=-14.85\times10^{-6}$\,e.m.u./mol \cite{Mendelsohn1970}).  These results thus indicate that there is little to no enhancement of the paramagnetic susceptibility due to ferromagnetic interactions.

\subsection{Magnetic Resonance}

\subsubsection{Electric Field Gradient}

\begin{table*}
\caption{\label{tab:EFGs}EFG parameters for the Ga(1), Ga(2) and La sites in LaNiGa$_2$ determined from NQR and NMR spectra. $(a,b,c)$ correspond to the unit cell axes shown in Fig. \ref{fig:Laspec}(a). $\nu_Q$ is defined as $\nu_{zz}\sqrt{1+ \eta^2/3}$, where $\nu_{zz}$ is the largest eigenvalue.  Computed values are from band structure calculations as described in the text.}
\begin{ruledtabular}
\begin{tabular}{lddddd}
\textrm{site} & \multicolumn{1}{r}{$\nu_{aa}$ (MHz)} & \multicolumn{1}{r}{$\nu_{bb}$ (\textrm{MHz})} & \multicolumn{1}{r}{$\nu_{cc}$ (\textrm{MHz})} & \multicolumn{1}{r}{$\nu_Q$ (\textrm{MHz})} & \multicolumn{1}{c}{$\eta$} \\ \hline
        Ga(1) measured & -2.61 \pm 0.01 & 3.86 \pm 0.01 & -1.25 \pm 0.01 & 3.94 \pm 0.01 & 0.35\pm 0.01 \\
        Ga(1) computed & -1.091 & 4.332 & -3.241 & 4.506 & 0.496 \\ \hline
        Ga(2) measured & -0.78 \pm 0.01 & -7.99 \pm 0.01 & 8.77 \pm 0.01 & 9.70 \pm 0.01 & 0.82 \pm 0.01 \\
        Ga(2) computed & -0.060 & -7.011 & 7.072 & 8.131 & 0.983 \\ \hline
        La measured & -0.86 \pm 0.01 & 1.65 \pm 0.01 & -0.80 \pm 0.01 & 1.71 \pm 0.01 & 0.04\pm0.01 \\
        La computed & -0.948 & 1.907 & -0.959 & 1.907 & 0.005 \\
\end{tabular}
\end{ruledtabular}
\end{table*}


Fig. \ref{fig:UnitCell}(b) shows the NQR spectrum measured at 4 K. There are several peaks, and it is not obvious a priori which transitions correspond to which site. In an applied field the $^{139}$La NMR spectrum (Fig. \ref{fig:Laspec} in the appendix) reveals seven transitions at frequencies split by $\nu_{\alpha\alpha}$. These splittings enable us to identify the EFG at the La site and hence the three peaks in blue shown in the NQR spectrum. The fitted values of the tensor elements are given in Table. \ref{tab:EFGs}.  The EFG vector (the direction corresponding to the largest eigenvalue of the EFG tensor) lies along the $b$-direction. However, the EFG asymmetry paramter is remarkably small: $\eta = 0.04\pm 0.01$. This indicates that the three La peaks in Fig. \ref{fig:UnitCell}(b) approximately correspond to the three transitions of $I_z = \pm1/2 \leftrightarrow \pm3/2$, $I_z = \pm3/2 \leftrightarrow \pm5/2$, and $I_z = \pm5/2 \leftrightarrow \pm7/2$ from low to high frequency peaks. The NQR frequency of the La is similar to that observed in LaNiC$_2$, which has a similar structure but is non-centrosymmetric \cite{Iwamoto1998}.

The remaining resonances in the NQR spectrum are due to the transition of $I_z = \pm1/2 \leftrightarrow \pm3/2$ at the two Ga sites in the unit cell [Ga(1) and Ga(2) in Fig.\ref{fig:UnitCell}(a)], which each have two spin 3/2 isotopes ($^{69}$Ga and $^{71}$Ga), giving us in total four transition peaks.  To identify the EFG at these sites, we performed NMR in an applied field as a function of angle, as discussed in the appendix.  The fitted values of the EFG are given in Table \ref{tab:EFGs}. There is a large asymmetry parameter $\eta$ for both sites, reflecting the orthorhombic nature of the local electronic environment.   The EFG vector for one of the two Ga sites lies along the $b$-axis, similar to the La site, however for the other site the EFG vector lies along the $c$-axis.

In order to discern the transitions for the two different sites and two isotopes, we turn to the DFT calculations, whose values are given in Table \ref{tab:EFGs}. For both the La and Ga(1) sites, the EFG vector lies along the $b$-axis, but for the Ga(2) site it lies along the $c$-axis, enabling us to assign the two Ga resonances. We find that the lower frequency peak with $^{69}\nu_Q = 3.94\pm 0.01$ MHz corresponds to the Ga(1) site, and the higher frequency peak with $^{69}\nu_Q = 9.70\pm 0.01$ MHz corresponds to the Ga(2) site.  The observed and theoretical values are within 20\% of each other.



The NQR spectrum in Fig. \ref{fig:UnitCell}(b) also reveals a smaller third resonance near 11 MHz.  The origin of this third resonance is unknown, although the NQR frequency is close to that of pure $^{69}$Ga metal \cite{Hwang1977}. It may also arise from an impurity phase, such as LaNiGa which is close to the composition of the flux and has been detected in powder x-ray diffraction~\cite{Badger2022}. The lower $^{71}$Ga resonance near 2.5 MHz was not obtained due to the limitations of the tuning range of the resonance tank circuit.

\subsubsection{Knight Shift and Spin-Lattice Relaxation Rate}

The temperature dependence of the magnetic Knight shift along the $b$ direction, $K_{bb}$, is shown in Fig. \ref{susceptibility}(b) for both the $^{139}$La and $^{69}$Ga(1) sites.  The shift is largely temperature-independent up to 100 K, and exhibits a small increase ($\sim 20\%$) between 100-300 K for both sites.  The shift arises due to hyperfine couplings between the nuclear spin and both the orbital and the Pauli spin components of the susceptibility \cite{CPSbook}. In general, the shift can be written as:
$K = K_{o} + K_{s}$, where $K_{o}=B_{D} \chi_{D}$ and $K_{s} = B_{P}\chi_{P}$ are orbital and spin contributions to the shift, and $B_{D,P}$ are the hyperfine coupling constants to these degrees of freedom. In materials where $K$ and $\chi$ vary with temperature, it is possible to extract $B_{P}$ by plotting $K$ versus $\chi$, but the temperature independence of these quantities precludes this approach in this case.  It is therefore not straightforward to determine what portion of $K_{bb}$ arises due to orbital versus spin contributions.

Fig. \ref{susceptibility}(c) displays the temperature dependence of the spin-lattice relaxation rate divided by temperature, $(T_1T)^{-1}$, measured with field along the $b$ direction. There is little to no temperature dependence evident for either site. There is a small increase in the La relaxation rate near 5 K, but this feature is not observed in the Ga and may be an artifact.

\section{Discussion}

Korringa behavior, or temperature-independence of $(T_1T)^{-1}$, is a hallmark of conductors and arises due to spin-flip scattering between the nuclear spins and the spins of the electrons at the Fermi surface \cite{Korringa1950}. For a single hyperfine coupling channel with an isotropic Fermi-contact type interaction, $T_1T K_s^2=S_0$ is a temperature independent constant, where $S_0 = (\gamma_e/\gamma_n)^2\hbar/(4\pi k_B)$, and  $\gamma_{e,n}$ is the gyromagnetic ratio of the electron/nucleus. Figure \ref{fig:Korringa} displays the Korringa ratio, $T_1T K_{bb}^2/S_0$,  for both the $^{139}$La and $^{69}$Ga(1) sites in LaNiGa$_2$ as a function of temperature, and compares this quantity with several other materials.  In principle we should use $K_{s}= K_{bb} - K_o$ rather than $K_{bb}$, however we are unable to independently measure $K_{o}$.  As a result, this  discrepancy likely gives rise to the fact that the ratio is different than unity for the $^{69}$Ga(1) and $^{139}$La sites.

In the presence of exchange enhancements of the conduction electron spin susceptibility, the Korringa ratio can deviate strongly from unity \cite{MoriyaT1formula,NarathWeaver}.  For the simplified case of a single, spherical Fermi surface, the ratio can be related to the Stoner enhancement factor.  In this case, a ratio greater than unity heralds ferromagnetic fluctuations, whereas a ratio less than unity indicates antiferromagnetic fluctuations.  This trend is evident in Fig. \ref{fig:Korringa} for several other materials known to exhibit either ferromagnetic order or antiferromagnetic fluctuations, including Sr$_2$RuO$_4$, which exhibits TRS breaking in the superconducting state \cite{Luke1998}, UPt$_3$, which exhibits non-unitary triplet superconductivity \cite{UPt3spintripletNMR}, and K$_2$Cr$_3$As$_3$, which exhibits chiral $p$-wave superconductivity \cite{Yang2021a}. {Although none of these materials exhibits a single band with a spherical Fermi surface, it is clear that LaNiGa$_2$ is significantly different, with} a temperature-independent ratio that is fairly close to unity for both the $^{139}$La and $^{69}$Ga sites.

Further evidence for a lack of correlations is provided by the fact that the measured EFG values are relatively close to those computed by the DFT band structure. In materials that exhibit strong correlations, the measured EFGs can differ significantly from those computed via band structure \cite{MeierEFG2002,Asadabadi2007}.  The calculations used here for LaNiGa$_2$ did not include any Coulomb repulsion terms, but still are within 20\% with the measured values.  This fact suggests that correlations are relatively small in this material.

In summary, we find that LaNiGa$_2$ does not exhibit any significant Stoner enhancement or evidence of enhanced spin fluctuations.  The unusual superconducting state, and TRS breaking below $T_c$, must therefore arise from the topological nature of the band structure, rather than from large electronic interactions that are believed to drive unconventional superconductivity in many strongly correlated systems. Our results motivate revisiting $\mu$SR studies on single crystals to better understand the TRS breaking.

\textit{Acknowledgment.} We acknowledge helpful discussions with W. Pickett and F. Ronning. Work at UC Davis was supported by the NSF under Grant No. DMR-2210613 and by the UC Laboratory Fees Research Program (LFR-20-653926). NQR measurements at LANL were performed with support from the UC Fees Research Program. ASB and MCY were supported by the NSF under Grant No. DMR-2323971. The XPS work was supported by the Alfred P.Sloan Foundation (FG-2019-12170).

\appendix

\renewcommand{\thefigure}{A\arabic{figure}}

\setcounter{figure}{0}
\begin{figure}[!h]
	\centering
	\includegraphics[scale=0.87]{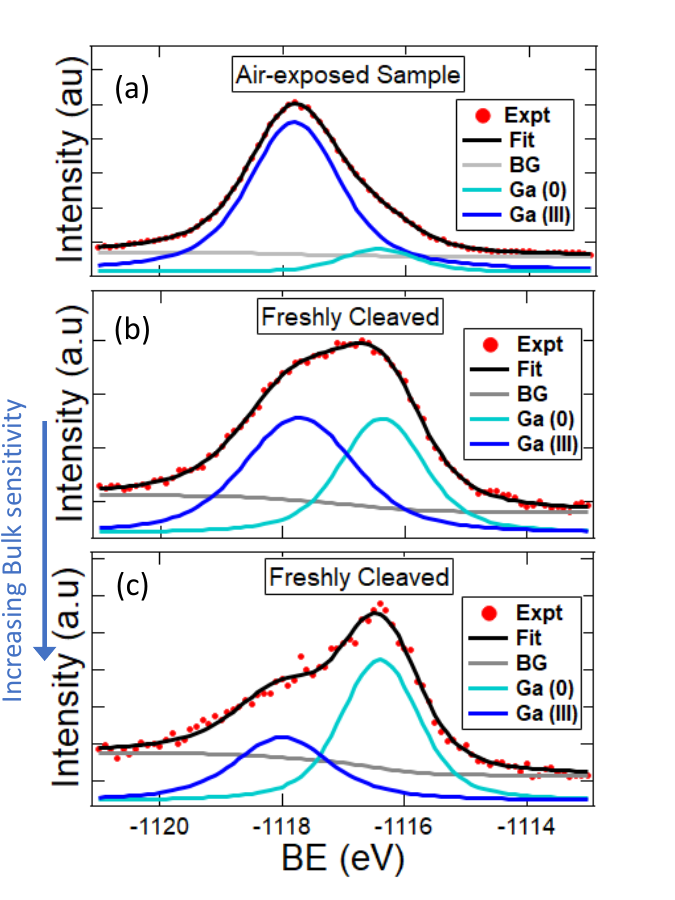}
	\caption{Fig 1: X-ray Photoemission Spectroscopy of Ga-$2p_{3/2}$ in LaNiGa$_2$. (a) Air-exposed spectrum;  (b) Spectrum of the same sample cleaved ex-situ before measurement; and  (c) Spectrum of the sample as in (b) but with Ag L-$\alpha$ source.
	\label{fig:XPS}}
\end{figure}

\section{XPS}

Fig. \ref{fig:XPS} displays a series of Ga $2p_{3/2}$ core level spectra under different conditions with different degrees of surface versus bulk sensitivity. The spectra were then fit with a Shirley background \cite{Shirley} and two Voigt lineshapes each corresponding to surface and bulk species. The binding energies across data sets (air-exposed vs freshly cleaved vs Ag source) were kept consistent. The binding energies were calibrated to a reference gold $4f$ spectrum. The Lorentzian width and Gaussian widths of the gold reference were found to be 0.385 eV and 0.339 eV, respectively.

\section{NMR}
\renewcommand{\thefigure}{B\arabic{figure}}

\setcounter{figure}{0}

\begin{figure}
\centering
\includegraphics[width = \linewidth]{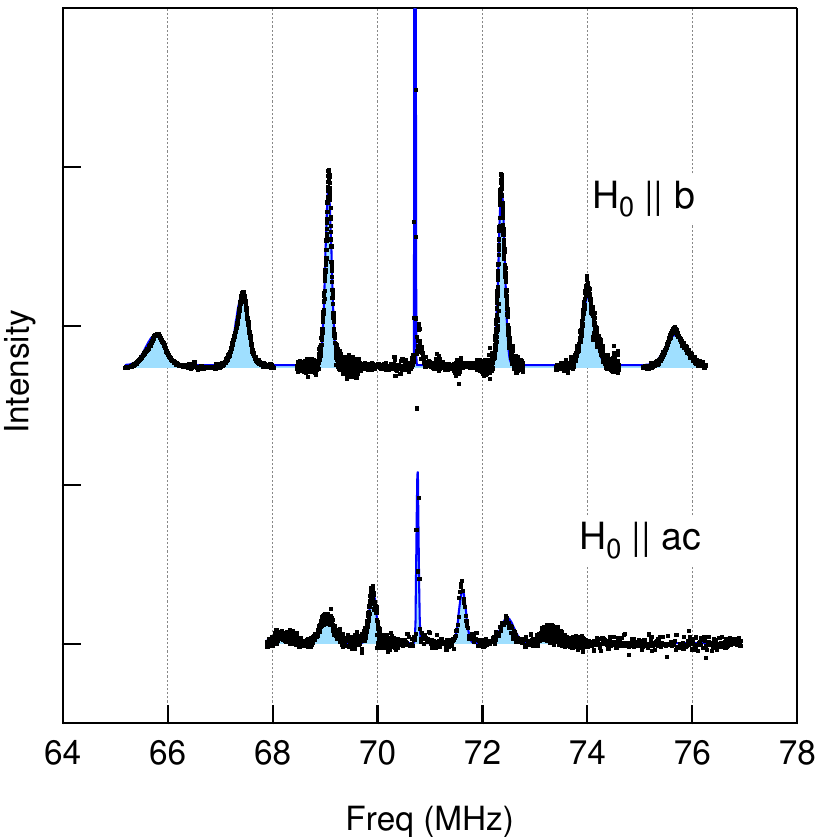}
\caption{$^{139}$La NMR spectra in LaNiGa$_2$ in an external magnetic field of $\mu_{0}H_0 = 11.7286$ T at 5 K for fields parallel and perpendicular to $b$. The filled regions show a fit to the spectra to an exact diagonalization of Eq. \ref{eqn:H}.}
\label{fig:Laspec}
\end{figure}

\begin{figure}
\centering
\includegraphics[width = \linewidth]{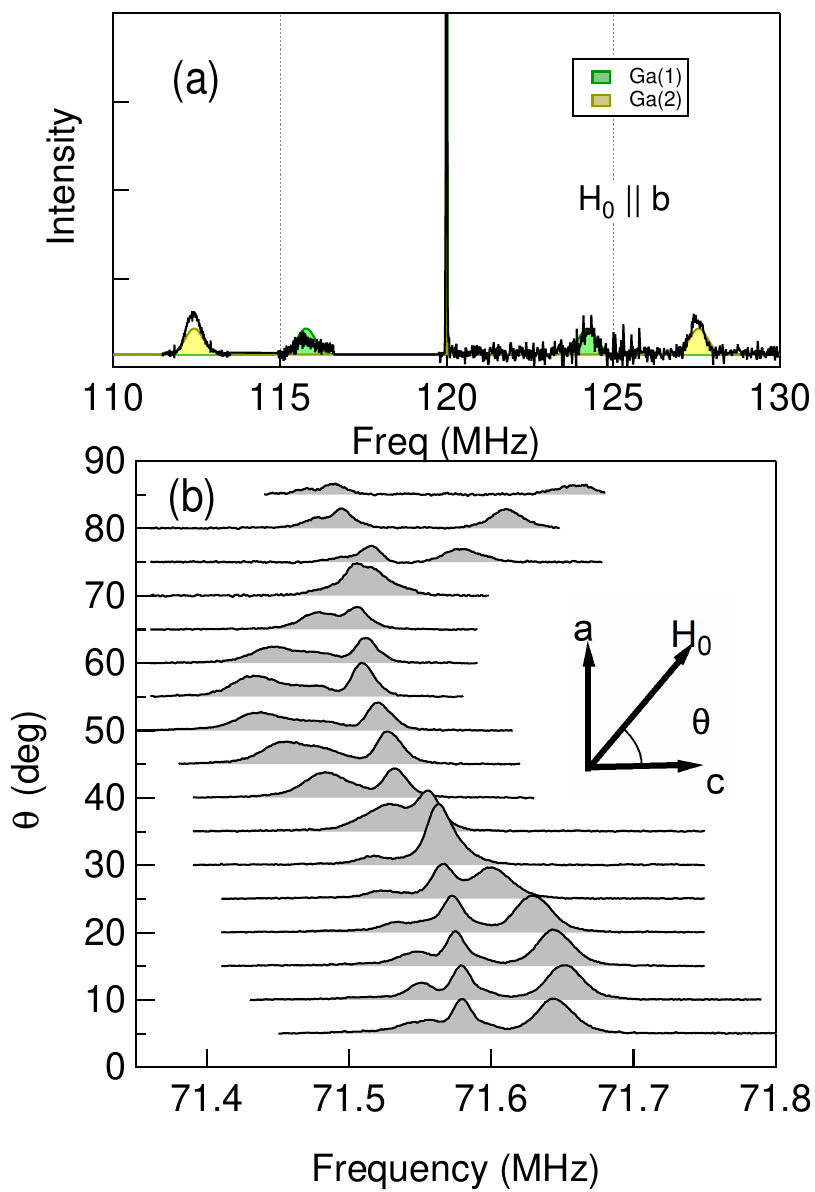}
\caption{(a) $^{69}$Ga spectrum in LaNiGa$_2$ for $\mathbf{H}_0\parallel b$ for $\mu_{0}H_0 = 11.7286$ T at 2.7 K. The filled regions show fits to the spectra to an exact diagonalization of Eq. \ref{eqn:H}.  (b)  Central transition of $^{69}$Ga as a function of field orientation in the $ac$-plane, where the angle, $\theta$, is measured relative to the $c$-axis, at 7.0 T and 2.7 K. }
\label{fig:waterfall}
\end{figure}

\begin{figure}[h]
\centering
\includegraphics[width = \linewidth]{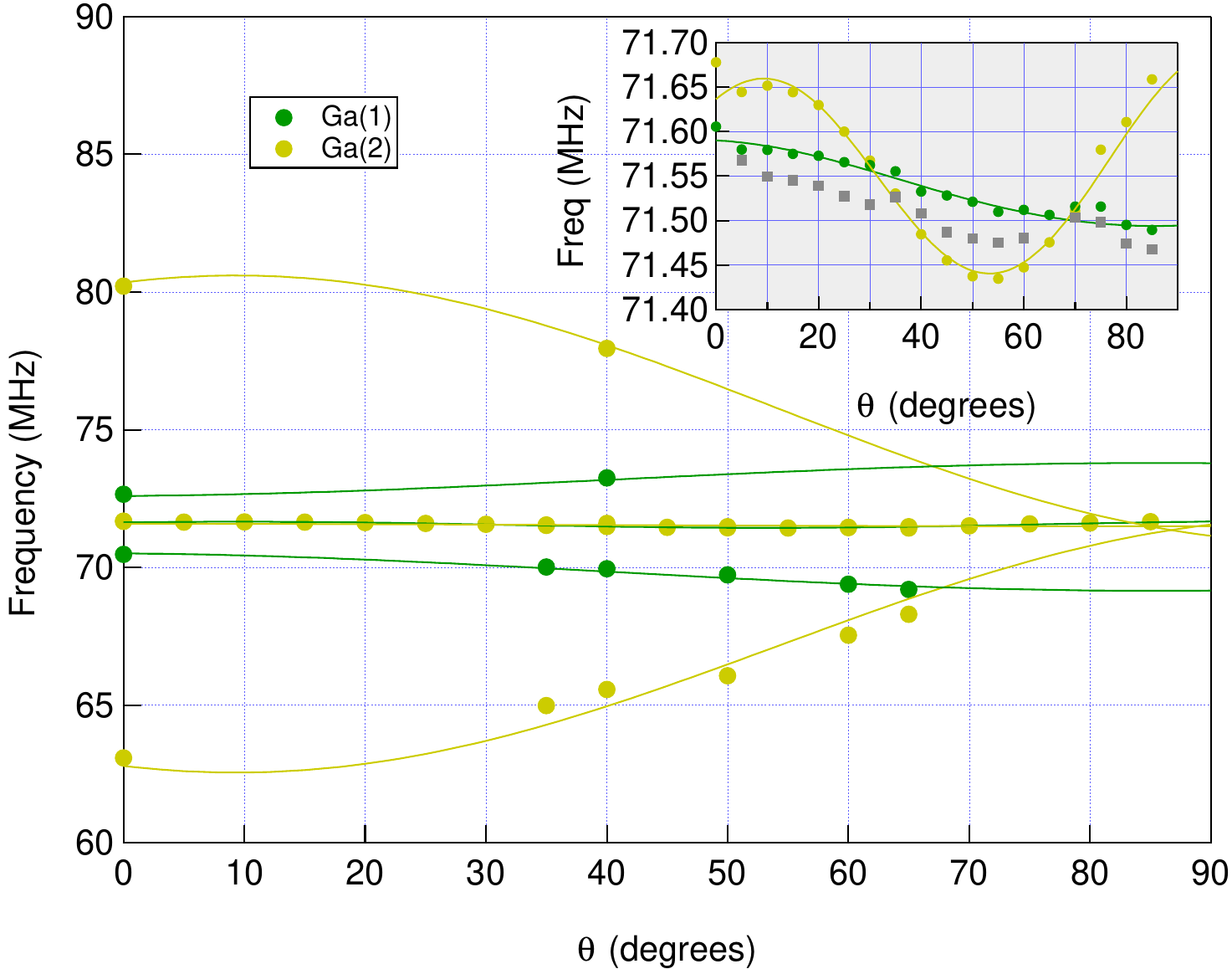}
\caption{Angular dependence of the $^{69}$Ga resonances as a function of $\theta$.  The solid lines are fits as described in the text.  The inset focuses on the central transition alone.  }
\label{fig:GaRotations}
\end{figure}

In a magnetic field, the La spectrum is split into seven resonances as shown in Fig. \ref{fig:Laspec}.  The Ga, on the other hand, is split into three resonances for each site, as shown for $^{69}$Ga for field along the $b$-axis  in Fig. \ref{fig:waterfall}(a). This spectrum shows two sets of quadrupolar satellites, and narrow overlapping central resonances.  To better determine the EFG tensor elements, we measured the spectrum as a function of field orientation in the $ac$ plane, as shown in Fig. \ref{fig:waterfall}(b) for the central transition.  There are three resonances visible, and their angular dependence is shown in Fig. \ref{fig:GaRotations}. Two of the peaks have roughly equal intensity, and there is a third peak at lower frequency with slightly lower intensity. The origin of this third peak is unknown, and we do not observe any associated quadrupolar satellite peaks. The angular dependence of the central and satellite peaks were globally fit for each site using perturbation theory to extract the EFG tensor elements, $\nu_{aa}$ and $\nu_{cc}$ in Eq. \ref{eqn:H}, and the fitted values are given in Table \ref{tab:EFGs}.

\clearpage

\bibliography{LaNiGa2Bibliography, LaNiGa2Mag, biblio}

\end{document}